\documentclass[aps]{revtex4}
\usepackage{graphicx}

\begin{document}
\vskip 20pt
\hbox to \hsize{\hfill SLAC-PUB-14274}
\vskip 12pt

\title{\vskip 20pt
Massive Degeneracy and Goldstone Bosons: A Challenge for the Light
Cone\footnote{This work was supported by the U.~S.~DOE, Contract No.~DE-AC02-76SF00515.}
\footnote{Talk given at Light Cone 2010 - LC2010,        June 14-18, 2010 , Valencia, Spain}
}
\author{Marvin Weinstein}
\affiliation{
       SLAC National Accelerator Laboratory,\\
       Stanford, CA, USA\\
       E-mail: niv@slac.stanford.edu}

\def\be{\begin{equation}}
\def\ee{\end{equation}}
\def\bea{\begin{eqnarray}}
\def\eea{\end{eqnarray}}
\def\bra#1{\left< #1 \right\vert}
\def\ket#1{\left\vert #1 \right>}
\def\braket#1#2{\left< #1 \vert #2 \right>}

\begin{abstract}
Wherein it is argued that the light front formalism has problems dealing with
Goldstone symmetries.  It is further argued that the notion that in hadron condensates
can explain Goldstone phenomena is false.
\end{abstract}
\maketitle

\section{A Definition}

You might well ask why I have decided to talk about the meaning
of the Goldstone theorem and physics on the Light
Front.  After all, while important, this is
not really a new subject.  The answer is simple.  I have chosen to give this
talk, rather than talk about new research, because at past Light Cone
meetings I have discovered that this subject is not as well
understood by all of the participants as it should be.  Therefore,
people have not understood the discussions that have taken place
between Stan Brodsky and me on the question of {\it in hadron condensates\/}
and the problems with the uniqueness of the Light Front vacuum state.
I have titled this talk "Massive Degeneracy and Goldstone Bosons:
A Challenge for the Light Cone" to emphasize the importance of this
issue for a proper understanding of PCAC and Current Algebra.

As it turns out, during my recent travels I found that my titles are not
always easily understood.  So I went back to the Wikipedia
for a definition of {\it degeneracy.\/} This is what I found:
\begin{itemize}
\item{\bf de-gen-er-a-cy [ di?jénn?r?ssee ] noun (plural de-gen-er-a-cies)}
\begin{description}
\item[Definition:]
\end{description}
\begin{enumerate}
\item bad behavior: immoral, depraved, or corrupt behavior, or an instance of this
\item worsened condition: a condition that is worse than normal or worse than before
\item worsening of condition: the process of becoming physically, morally, or mentally worse
\item quantum physics states of equal energy: the condition of two or more quantum states having
the same energy.
\end{enumerate}
\end{itemize}
I would love to discuss the first three topics, but alas, I will
be discussing the fourth in this talk.  The point that I want to make in the time
allotted to me, is that the successes of PCAC and Current Algebra require us to
conclude that in our world the hadron sector is very close to a theory where
$ SU(3) \times SU(3)$ is realized as a Goldstone (other people say spontaneously
broken) symmetry.  A corollary of this, is that in the limit
of an exact Goldstone symmetry, the ground state of the theory is enormously
degenerate and this feature of QCD is not apparent
in the Light Front formulation of the theory.  {\it This is the feature of QCD
that has to be better understood.\/}

\section{A Parable}

What would a general talk be without a parable ? Nothing!  So, having said that,
I will begin with a modified version of a parable that I believe
I first heard from Sydney Coleman.

Once upon a time in a universe far, far smaller than ours there
lived the famous savant Doctorus E, who was an expert on practically
everything and worked at the famous Crystallus U.
One day as the Doctorus was deep in thought a student
interrupted with a strange observation.  He said, "Doctorus, I have
just discovered an amazing thing!  The world is translation invariant!"
Never at a loss for a response the famous savant replied "Dummkopff, everybody knows
that!
\begin{figure}[h!]
 \vbox{
 \hbox to \hsize{\hbox{\hss
  \includegraphics[width=2.15in]{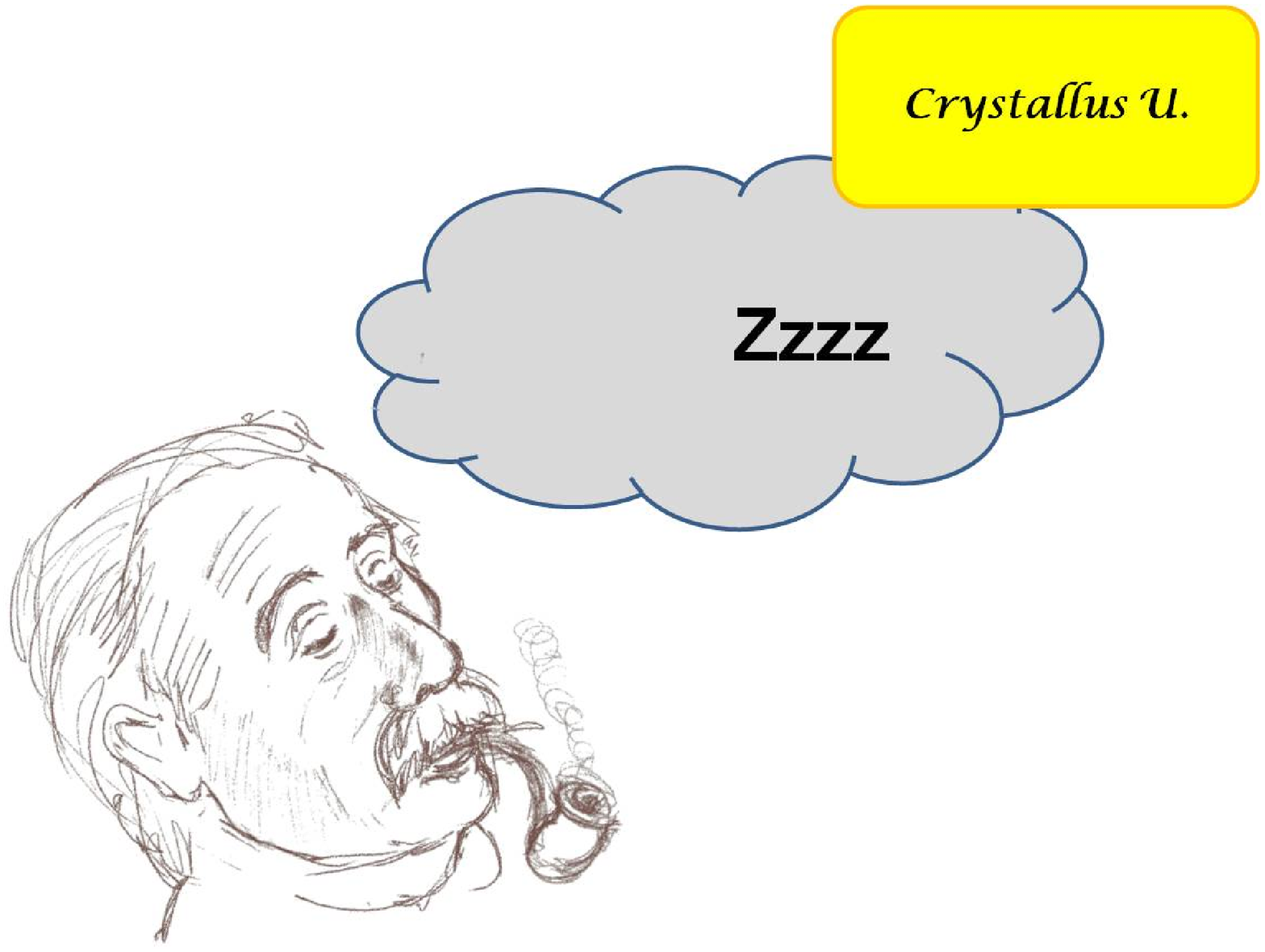}
  \hss}
  \quad
  \hbox{\hss
  \includegraphics[width=2.15in]{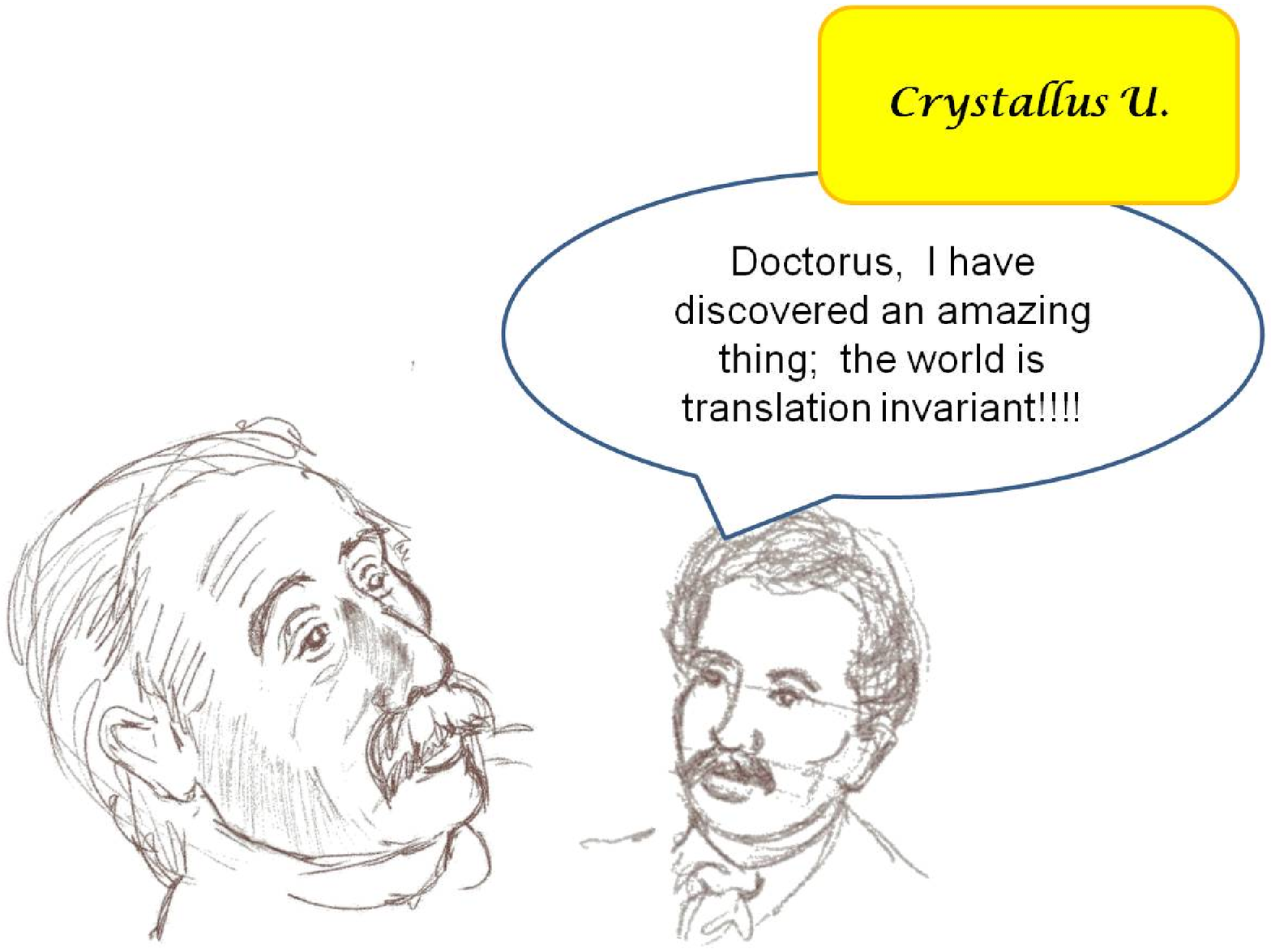}
  \hss}
  }
  \hbox to \hsize{\hss
   \includegraphics[width=2.15in]{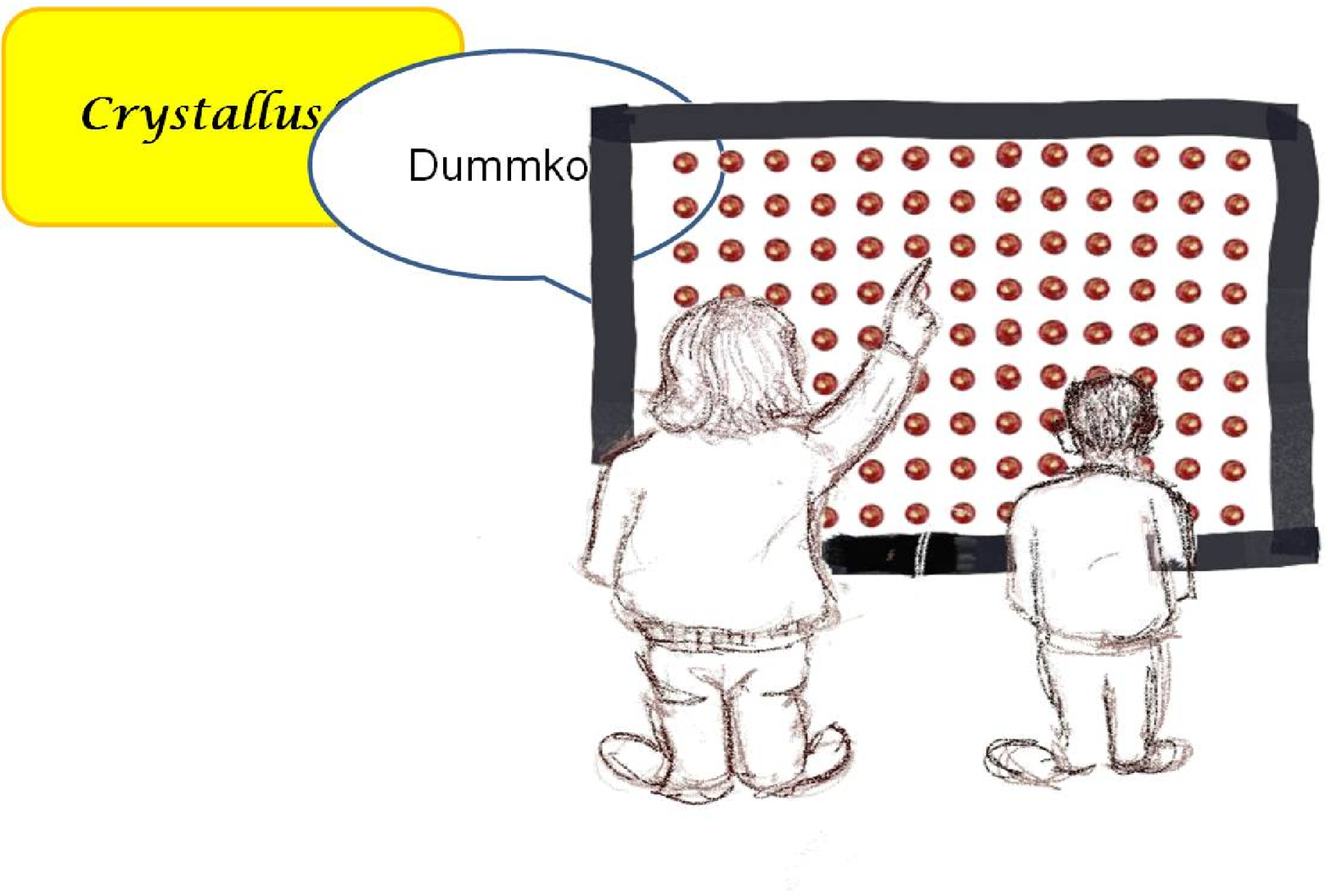}
  \hss}
  }\vskip -30pt
  \caption{Read the pictures left to right and top to bottom}
\label{doctorus}
\end{figure}
Come here and look out the hyper-viewer at what can be seen in
the sky.  Obviously if you move the entire world over by one grelp (an astronomical unit
in Crystallus) everything looks the same.  We have known that forever".

"But no Doctorus", said the student,"that is not what I meant!  I mean
that the laws of physics don't change even if we translate the world
by an arbitrarily small amount".  To which the savant responded "How could
you know that ?  It would take an infinite amount of energy to move
the whole world by an arbitrarily small amount!  How could we test this
idea ?"

"That's the neat thing Doctorus", said the student.  You don't have
to move the whole world to find out if there is, what I call, a {\it
hidden symmetry\/}.  All you have to do is see how much energy it
takes to excite an arbitrarily long wavelength excitation.  The
consequence of the hidden symmetry is that this energy will go to
zero as the wavelength goes to infinity.  In addition, these long
wavelength excitations satisfy sum rules constraining their interactions
with imperfections"

"Hmm", said the savant, "let me think about this."

As it turned out the young student was correct.  In fact, parable aside,
translation invariance (by arbitrarily small translations) is realized
as a {\it hidden symmetry\/} in a crystal.  The excitations are called
{\it phonons\/} and these phonons have no mass (i.e., their energy
as a function of momentum goes to zero as the momentum vanishes).

\section{Examples of Goldstone Symmetries}

Of course, I wouldn't be talking about this if the only example of a
hidden symmetry was phonons in a crystal.  In fact, condensed matter
physics is replete with examples of systems that exhibit this phenomenon.

For example, ferromagnets are objects that exhibit spontaneous magnetization;
i.e., in these systems magnetic moments of atoms align with one another
to produce an observable, persistent magnetic field.  Since this magnetic field
points in a definite direction, it follows that the rotational invariance of
the full theory is {\it hidden\/} from us.  The massless excitations (or gapless
excitations of this system are called {\it magnons\/}).  Since the
manetization could point in any direction, it follows that the ground state
of this system is infinitely degenerate.  Similarly, there are anti-ferromagnets,
\begin{figure}[h!]
 \hbox to \hsize{\hbox{\hss
  \includegraphics[width=2.25in]{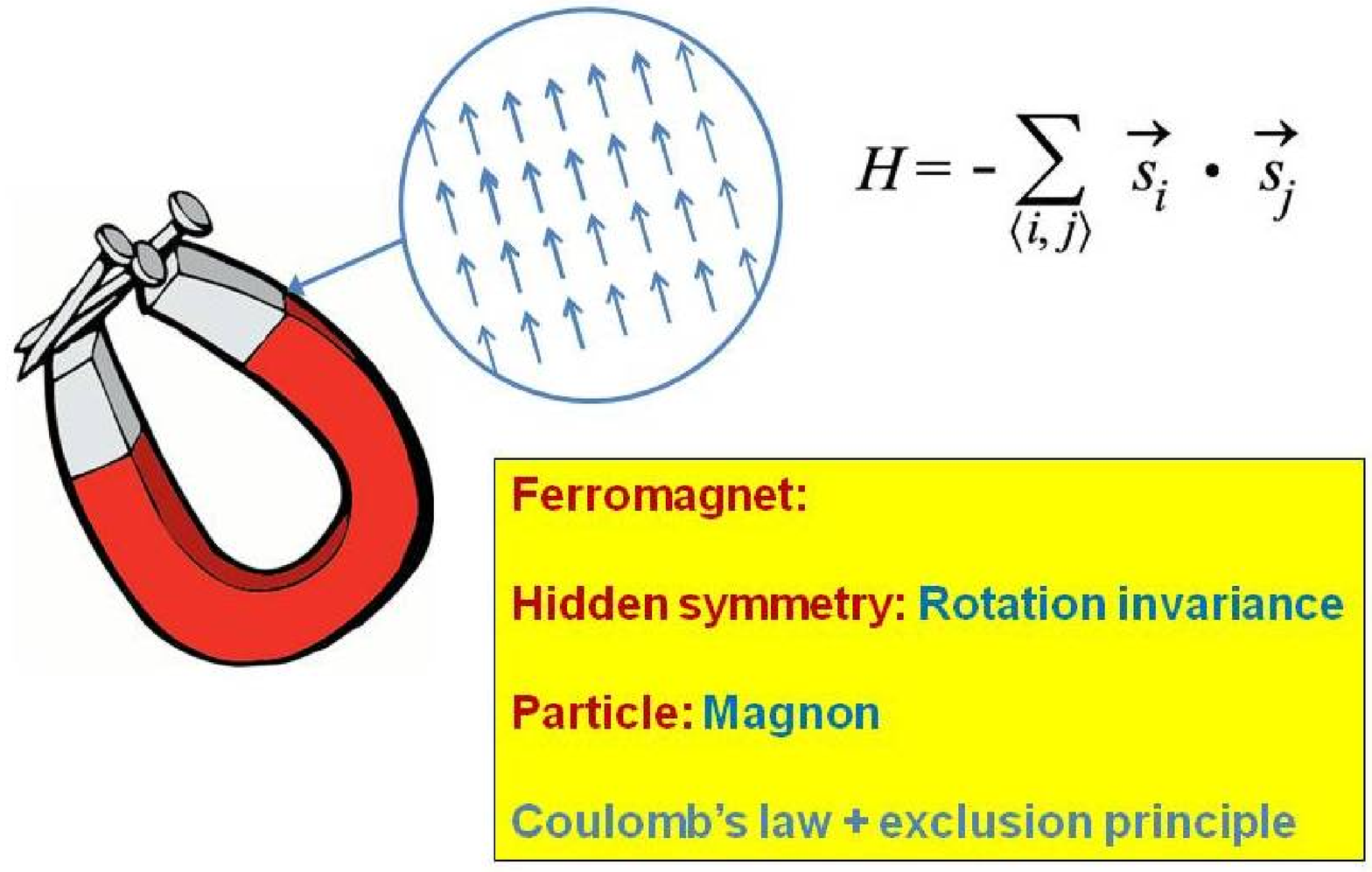}
  \hss}
  \quad
  \hbox{\hss
  \includegraphics[width=2.25in]{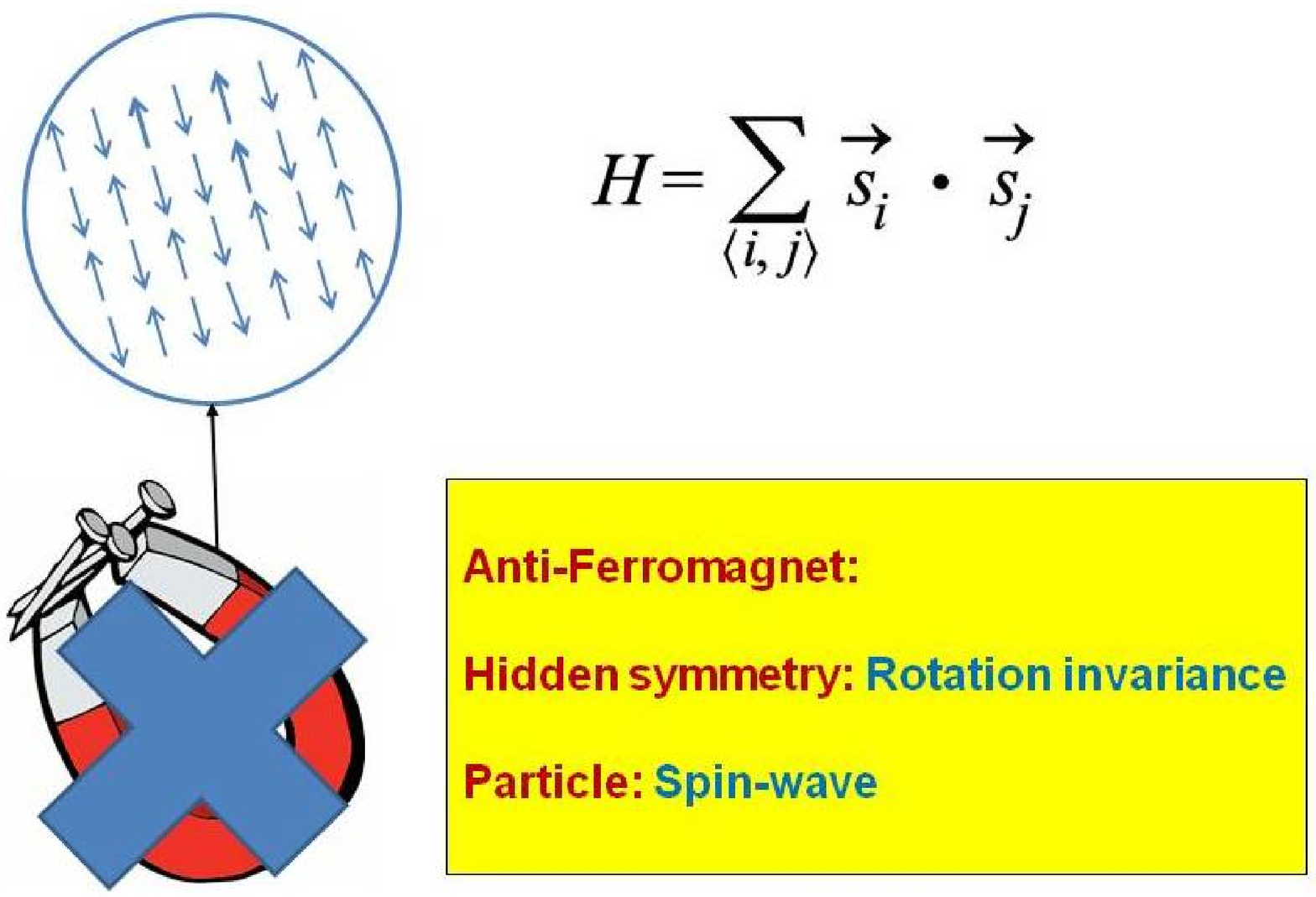}
  \hss}
  }
  \caption{Both the ferromagnet and anti-ferromagnet have rotational invariance
  as a hidden symmetry.}
\label{doctorustwo}
\end{figure}
where the spins anti-align, but point in a definite direction.(High temperature
superconductors exhibit this behavior when they are under-doped.)
Once again the hidden symmetry is rotational invariance and the massless
excitations are called {\it spin-waves\/}.

Finally, not to be outdone by the world of condensed matter physics, particle physics
also has its share of hidden (or Goldstone) symmetries.  For the purposes of this
discussion I will avoid theories that have the additional complication of the
Higgs phenomenon taking place and will concentrate on QCD.  It is an old
story that the only consistent explanation of the Goldberger-Treiman relation,
Adler-Weissberger calculation of the $g_A/g_V$ sum-rule, the sum rule for the
squares of the masses of the pseudoscalar mesons, Weinberg's formula for the
$\pi-\pi$ scattering lengths, Dashen-Weinstein theorem on the slope of the
form factors in $K l_3$ decay, etc., is that QCD is close to a theory in which
$SU(2) \times SU(2)$ (and in fact $SU(3) \times SU(3)$) is an exact but {\it hidden\/},
(or Goldstone) symmetry.  Moreover, the only source of symmetry breaking
are the quark mass terms in the Hamiltonian. In other words, if quark masses were
zero, then this symmetry would be exact and the $\pi$, $K$ and $\eta$ meson would
all have zero mass.  They would be, like the phonons, magnons or spin-waves, the
massless excitations associated with the hidden symmetry. By the way, one reason
I like calling this sort of symmetry a hidden or Goldstone symmetry, rather
than the more popular {\it spontaneously broken\/} symmetry is that it
avoids having to talk about a "really broken, spontaneously broken, symmetry"
when quark masses are non-zero.

The key point I want to make in the rest of this talk is that it is genrally true,
as is obvious in the case of the condensed matter examples I talked about, that
in order for there to be Goldstone bosons (these massless excitations) there has
to be something to {\it wiggle\/}.

\section{Formalities}

Since I see friends in the audience with a bit of a formal bent, I will spend
a few moments giving some formal insight into the meaning of the Goldstone
theorem and what it says about there having to be stuff that wiggles.
A more extensive discussion of some of these issues can be found in my Heidelberg
lectures\cite{Weinstein:1971ed} .

The story begins with that perennial favorite, Noether's theorem.
You all know the theorem, you learned it in grade school.  It says, that
if a Lagrangian has a continuous symmetry then there exists a locally
conserved current associated with that symmetry; i.e.,
\be
    \partial_{\mu}\,j^{\,\mu} = 0 .
\label{Neother}
\ee
The usual follow up to the proof of this theorem is the observation
that as a consequence of current conservation there is an associated
conserved charge
\be
    Q = \int\,d^{\,3}x \,j^{\,0}(x) .
\label{Qexact}
\ee
It is usually argued that a consequence of the current conservation equation that the
time derivative of this charge vanishes
\be
  \partial_0\, Q = \int\,d^{\,3}x \  \, \partial_0 \,j^{\,0}(x) = - \int\,d^{\,3}x\  \,\vec{\nabla}\cdot
  \vec{j}(\vec{x}) = 0 ,
\label{dQdt}
\ee
at least if surface terms can be neglected.  Then, so the story goes,
we have a time independent Hermitian operator that can be exponentiated
to provide a unitary representation of the symmetry group.
It is here that the story becomes more complicated.  As with all things
in physics there is often a {\it gotcha\/}.

To better understand what the gotcha is, define the operator
\be
    Q_{\Omega} = \int_\Omega\,d^{\,3}x \,j^{\,0}(x) ,
\label{Qomega}
\ee
where the integration is over a finite three volume $\Omega$.  The
good thing about his operator is that it exists and the local
nature of the commmutator of the current with local fields guarantees
that
\be
    \lim_{\Omega \rightarrow \infty} \left[ \, Q_\Omega\, , \phi(\vec{x}) \right]
\label{commutatorwithfield}
\ee
exists, since once $\Omega$ is larger than the intersection of either the past
or future light cone of the point $\vec{x}$ with the surface of integration
in Eq.\ref{commutatorwithfield}.  This observation tells us that by taking the
limit $\Omega \rightarrow \infty$ in all commutators of $Q_\Omega$ with all
local observables we obtain an automorphism of the space of local observables:
the question is whether this is an {\it inner automorphism\/}?  In other words,
is the a Hermitian operator $Q$ defined on the Hilbert space that generates the
same automorphism.  If so, it can be exponentiated.  Furthermore, if the conserved
currents close to the algebra of a compact Lie group, then so will the conserved
Hermitian charges and therefore they will generate a unitary representation of
the Lie group on the space of physical states.  In that case we say that the
conserved currents are realized as a {\it Wigner symmetry\/}.  The hallmark of a
Wigner symmetry is that the states are grouped into finite dimensional representations
of the Lie group (since it is compact) and they all have the same mass; i.e., they
are degenerate.  Also, using the Wigner-Eckart theorem, we are able to relate
matrix elements of operators that transform as irreducible representations of the
group, to one another.  Hence, we get relations between coupling constants,
transition matrix elements, etc.

Suppose, however, that the automorphism defined by the conserved currents is
not inner?  This can only be the case if the limit of the operator $Q_\Omega$
fails to exist as $\Omega \rightarrow \infty$.  This will happen if there is
a massless particle coupled by the current to the vacuum state.  When this
happens the existence of the conserved currents implies that the system
has a non-trivial symmetry, but this symmetry is no longer realized by
having the states of the theory bundled into nice finite dimensional
irreducible representations of the Lie group.  Rather, the consequences
of the symmetry are exact low-energy theorems controlling the low energy
behavior of the massless particle that is coupled by the conserved
currents to the vacuum.  Such a symmetry is not immediately obvious to
us and for that reason I will adopt Coleman's terminology and refer to
it as a {\it hidden\/} or {\it Goldstone\/} symmetry.  This is the sort
of symmetry realized by the conserved axial vector currents in QCD in the
limit of vanishing quark masses.  The Goldstone bosons, i.e. the massless
particles coupled to the vacuum state by these currents, are the
$\pi, K $ and $\eta$ mesons.  Some of the consequences of this symmetry
are:  the Goldberger-Treiman relation, the Adler-Weissberger realtion,
the PCAC self-consistency conditions, the Dashen-Weinstein theorem on
the form factors in $K\,l_3$ decay, and Weinberg's theorem on the behavior
of low energy $\pi - \pi$ scattering.  This list is by no means
complete, I give it only to convince you that there is a great deal of
experimental evidence that points to the fact that the axial current
part of chiral $SU(3) \times SU(3)$ is realized as a hidden or Goldstone
symmetry, whereas the symmetry generated by the vector currents is of the
Wigner type.

Of course, since this talk is about massive degeneracies, I should
explicitly point out that the existence of the massless particle
created by this current means that the lowest energy state of the
theory is enormously degenerate.  This is because we can add any
number of zero momentum massless particles to the vacuum without
increasing the energy.  It is the fact that {\it the light-front formalism
insists that the vacuum is the unique lowest energy
state\/} that makes reconciling the light
front treatment of QCD with the real world so problematic.  I would also
like to point out that any argument that says {\it in hadron condensates\/}
can explain the Goldstone boson phenomenon is simply incorrect, in that
it cannot explain this huge vacuum degeneracy.

\subsection{Finite Volume Considerations}

I now have to say a few words about what happens if I make the spatial
volume finite, instead of infinite.  Why do I feel compelled to do
this?  Because all non-perturbative approaches to dealing with
QCD involve beginning with a system in a finite volume and then
taking the volume to infinity, and when the volume is finite, the
ground state of the theory is unique.  Where then is the massive
degeneracy I spoke of?

To clarify this issue, let us return to the example of the
ferromagnet.  Imagine we make a state that is the tensor product
of essentially the same norm one spin state on each lattice site,
and assume that this state is constructed so that the expectation
value of the spin (or magnetization) points in a definite direction.
This product state is a contender for the infinite volume magnetized state.

If one now applies a rotation to the spins on each site one obtains a new state
for which the magnetization points in another direction.  Now, because
the volume is finite the overlap of two such states is given by
\bea
    \ket{\Psi} = \prod_i^N \ket{\psi_i}\quad {\rm and} \quad
    \ket{\Phi} = \prod_i^N \ket{\phi_i} \\
    \braket{\Psi}{\Phi} = \prod_i^N\, \braket{\psi_i}{\phi_i} .
\eea
Note, since two non-aligned states of unit length have an overlap whose
magnitude is less than one; i.e.,
$\vert\vert \braket{\psi_i}{\phi_i} \vert\vert < 1$, it follows that
the overlap of the finite volume states, $\Psi$ and $\Phi$ goes to zero
exponentially as that number to the power $N$.
Furthermore, for a spin-spin Hamiltonian it is clear that
\bea
    \bra{\Psi} \,H\,\ket{\Psi} = \bra{\Phi}\,H\,\ket{\Phi} ,\\
    \bra{\Psi}\, H\,\ket{\Phi}\ \approx \ X^N \ \rightarrow\  0 ,
\eea
where $X$ is some number less than unity.  Since the different
states are not orthogonal (but the are unit length) one can
use them to form an orthonormal basis and diagonalize the
Hamiltonian truncated to this space of states.  These will be
the correct lowest lying eigenstates in the limit of large $N$
and they will be split by an amount that goes to zero as
$N \rightarrow \infty$.  It is the fact that these states
become split by exponentially small amounts as the number of
sites gets large that explains how a finite size ferromagnet
seems to form.  Clearly, the state in which the ferromagnet
points in a definite direction is a linear combination of
the eigenstates we constructed.  Since the splitting between
these states is so small, turning on a small magnetic field
will put the system into this {\it magnetized\/} state.  If this field is then
turned off adiabatically the different eigenstates will evolve in time
with slightly different phases, due to their energy differences.
However since these energy differences are exponentially small,
it will take an exponentially long time to see the magnetization
vanish.

The key point of all of this, is not why we can see
ferromagnets that have a finite volume.  Rather it is that
that the signal of the infinite degeneracy of the infinite
volume limit, is an enormous number of nearly
degenerate states whose number grows rapidly with increasing
volume.  This enormous degeneracy is what a light front calculation,
done in finite volume, should see. This is what, to the
best of my knowledge, isn't apparent yet.  The statement that
the virtue of the light front is that the vacuum state is empty
(and unique) is, in the case of spontaneous symmetry breaking,
a problem, not a virtue.

\section{What Happens in the Instant Formalism ?}

Having criticized the light front approach because it
doesn't make it easy to address Goldstone symmetries, it would be
remiss of me to not argue that this problem is less
difficult in the instant formalism.  I will now
contend that for a non-Abelian gauge theory, such as QCD,
the formation of Goldstone bosons is an inescapable property
of the strong coupling limit of the theory.

\subsection{The Schwinger Model}

Given the limitations of time, I will begin with a very
short discussion of the $1+1$-dimensional Schwinger model, because
it exhibits most of the physics I wish to discuss. After that I
will make an even briefer foray into QCD.  The point of this,
as I have already said, is to show that for these
theories it is trivial to argue that the strong coupling limit
of the theory explains why the vacuum state can be very degenerate
and support the existence of Goldstone bosons.

We begin with the formulation of the lattice version
of the Schwinger model in $A_0 = 0$ gauge.  The setting for
the model is a $1$-dimensional lattice whose sites are labelled
by the integer $j$.  The fermions in this model live on the sites
and are represented by the two-component fermion field $\psi_j$ and
its conjugate $\psi_j^\dag$.  The Abelian gauge field of the model,
$A_j$, lives on the link joining the pair of lattice sites $j$ and $j+1$,
The conjugate variable to $A_j$ is the electric field variable $E_j = \dot{A_j}$ and
they satisfy the canonical commutation relations $[ A_j, E_{j'}] =i\,\delta_{j,j'}$.
Since the variable $A_0$ appearing in the Schwinger model Lagrangian, it follows
that the Maxwell equation coming from varying the Lagrangian
with respect to $A_0$ will not be an equation of motion.
With these definitions, if we follow the usual prescription, we
construct the Hamiltonian of the generic form of the lattice
Schwinger model:
\be
H = H_E + H_f,
\ee
where
\begin{eqnarray}
H_E &=& \frac {g^2}{2} \sum_{n} E_n^2  \nonumber\\
H_f &=& \sum_{n,n^{'}} (\psi^{\dag}_n)^{\alpha} K(n-n^{'})_{\alpha \beta}
\,e^{-i\,\sum_{j=n}^{n^{'}-1} A_j} (\psi_{n^{'}})^{\beta} ,
\end{eqnarray}
where the kinetic term $K(n-n^{'})_{\alpha\beta}$ is a two-by-two matrix for each value of
$n-n^{'}$, the fermion fields satisfy the anti-commutation relations
\be
    \left\{ (\psi^{\dag}_n)^{\alpha} , (\psi_{n^{'}})^{\beta} \right\} =
\delta_{n,n'}\,\delta_{\alpha,\beta}.
\ee
The link fields satisfy the harmonic oscillator commutation relations given above.

Now in any number of dimensions, the missing Maxwell equation is
just the Gauss law.  In one dimension this law takes the particularly simple
and suggestive form
\be
    G_j = E_{j+1} - E_j - \psi_j^\dag\,\psi_j = 0 .
\ee
The important fact is that although this equation is not one of
the Euler-Lagrange, or Heisenberg equations of the theory, it follows
from the specific form of the Hamiltonian that all of the
operators $G_j$ commute with the Hamiltonian; i.e.,
\be
   \left[ G_j, H \right] = 0\quad \forall j .
\ee
From this it follows that, although the theory contains states
that do not satisfy the Gauss law, we are free to restrict
ourselves to the subspace of states which do, since the Hamiltonian
will never take us out of this subspace.  Furthermore, since
the operators $G_j$ are precisely the operators that generate
local time-like gauge-transformations, all gauge invariant operators
must also commute with the $G_j$'s.  From this discussion we see that
by beginning in $A_0 = 0$ gauge we over quantize the theory, in that
canonical manipulations can create more states than we wish; however,
thanks to gauge invariance we can select a subspace of states that
gives us the theory we are interested in.

Obviously, since we have the choice of how to choose our basis states,
we see that satisfying the Gauss law will be most easily done if we
work in a basis in which the operators $E_j$ and $\rho_j = \psi_j^\dag\,\psi_j$
are diagonal.  Since we are dealing with fermions, we have the possibility
of having four possible fermion states: these correspond to the
state having no particles, one particle of charge $-1$, or one anti-particle
of charge $+1$, or finally, one particle and one anti-particle on a single
lattice site.  Note, if we impose the Gauss condition then specifying the
charges on each site completely (up to a constant background field which
we will take to be zero) specifies the state.  Note, since the
Hamiltonian only contains the operators $e^{i\,A_j}$, it can only couple
together states in which the electric field changes in absolute value
by one unit.

At this juncture I am in a position to keep my promise and argue that
in the limit $g^2 \rightarrow \infty$ the Schwinger model has an infinitely
degenerate ground-state.  Moreover, I will argue that for very large, but
finite $g^2$ the theory has a Goldstone boson that, due to the anomaly,
fails to appear as we take the limit $g^2 \rightarrow 0$.

I begin by noting the the term $\frac{g^2}{2} \,\sum_j \,E_j^2$ means
that when $g^2 \gg 1$ having any flux (i.e. any non-vanishing value
for $E_j$) costs a lot of energy.  Thus, the ground state of the theory
in this limit must be a state where $E_j = 0$ for all links.  By the
Gauss condition this means that the charge on each site must vanish.
However, we have already seen that there are two possible zero charge
states for each site.  Thus we see that in the large $g^2$ limit,
for a lattice with $V$ sites, there will be $2^V$ degenerate states with
energy zero.  All other states will have infinite energy.  Now, if we
take $g^2$ large but finite, then we observe that the kinetic term
can separate a pair and create a particle and anti-particle on different
sites joined by a unit of flux.  Since this is a high-energy state the
energy denominator appearing in perturbation theory is large and so
we are invited to treat the effects of the kinetic term on the ground
state by second order perturbation theory.  However, since the ground
is so degenerate, we must do degenerate perturbation theory, since the
different degenerate states are mixed by the kinetic term.  This leads
us to an effective Hamiltonian which is our friend the Heisenberg
anti-ferromagnet, and as I already said, this theory has a symmetry
that is realized in the Goldstone mode.  Of course, there are no
little magnetic spins in this case, rather the role of spin has to
do with the chiral charge of the state which, for each site, is the sum of the
particle and anti-particle number on that site minus one.

Time doesn't permit me to discuss this model further, especially the
interesting story of what happens to the Goldstone mode in the
continuum (i.e. $g^2 \rightarrow 0$) limit.  Since what happens in
this case is specific to the anomaly in the axial current, it presumably
is not relevant to the case of the octet of axial currents in QCD.
I refer you to
my paper with Kirill Melnikov\cite{Melnikov:2000cc} on the subject for all of
the details.

\subsection{What About QCD ?}

At this point I can only give the briefest summary of what happens in QCD.
If one works in the corresponding version of $A_0 = 0$ gauge, the
story parallels that of the Schwinger model.  Once again, gauge
invariance requires that in the large coupling limit the color
charge on every site must vanish in order to avoid having non-vanishing
flux on any link.  Thus, each site can have as many $q \bar{q}$ or
$ q q q$ states as are allowed by the exclusion principle.  If we simply
focus on the possible mesons that there are a large number of zero
energy states on each site. As in the case of the Schwinger model
we do degenerate second order perturbation theory to understand
what happens when we turn on the kinetic terms.

The result is that we obtain a frustrated $SU(12) \times
SU(12)$ anti-ferromagnet.  The frustration is due to the presence
of next nearest neighbor hopping terms. The same terms break the
global symmetry to chiral $SU(3) \times SU(3)$.  It is straightforward
to show that, in this system, the vector $SU(3)$ symmetry is realized
in Wigner fashion, meaning that there are degenerate $SU(3)$ multiplets
of particles with non-vanishing mass, but the axial part of the
symmetry is realized in Goldstone mode. The massless multiplet of mesons
are the $\pi$, $K$ and $\eta$ mesons.  Another bonus is that in this
limit we get the good predictions of the ratio of magnetic moments
obtained in the old $SU(6)$ symmetry scheme, but none of the bad
predictions.  A complete discussion of this approach appears in
paper by myself, Sid Drell, Helen Quinn and Ben
Svetitsky\cite{Svetitsky:1980pa}-\cite{Weinstein:1980jk}.

\section{The Challenge}

This talk can be summarized as follows:
\begin{itemize}
\item Exact symmetries can be realized in Wigner or Goldstone mode.
\item When a symmetry is realized in Wigner mode the states of the theory
form degenerate irreducible representations of the symmetry group and the
lowest energy state is unique.
\item When a symmetry is realized in Goldstone mode the lowest energy state
of the theory is infinitely degenerate, the states of the theory do not
form irreducible representations of the symmetry group and there are massless
particles coupled by the conserved currents to any one of the possible
ground states.
\item In finite volume the signal of a Goldstone realization of a symmetry
is that the number of nearly degenerate states grows rapidly with increasing
volume and the gap between these states shrinks exponentially with the volume.
\item The existence of a {\it condensate\/} such as the magnetization, for a
ferromagnet, or the staggered magnetization for an anti-ferromagnet, signals
a Goldstone symmetry.  This is because this condensate transforms non-trivially
under the symmetry transformations and so its existence implies the ground state
isn't unique.
\item PCAC means that the pion, kaon and eta are would be Goldstone bosons of the
theory where the quark masses are set to zero.  This interpretation is
overwhelmingly supported by experimental data.  This means that these particles
are really the wiggling of the order parameter or condensate.
\item Finally, in order for the Goldstone particle to exist there has to be
something to wiggle every place where the particle can exist. {\it  This means that
the condensate that is the order parameter for this Goldstone symmetry
cannot be confined to the interior of hadrons.}
\end{itemize}

Thus, to reiterate, the challenge for the Light Front is to show how the formalism gives rise to
this sort of pattern of degeneracy when the physical volume of space becomes large.

\end{document}